\documentclass[12pt,thmsa,a4paper]{article}
\usepackage{sw20aip}

\input tcilatex
\QQQ{Language}{
American English
}

\begin{document}

\author{Hans J. Wospakrik\thanks{%
email: hansjwospakrik@cyberlib.itb.ac.id}  and Freddy P. Zen\thanks{%
email: fpzen@bdg.centrin.net.id} \\
\\
Theoretical Physics Laboratory, \\
Department of Physics, Institute of Technology Bandung,\\
Jalan Ganesha 10, Bandung 40132, Indonesia}
\title{Inhomogeneous Burgers Equation and the Feynman-Kac Path Integral }
\date{}
\maketitle

\begin{abstract}
By linearizing the inhomogeneous Burgers equation through the Hopf-Cole
transformation, we formulate the solution of the initial value problem of
the corresponding linear heat type equation using the Feynman-Kac path
integral formalism. For illustration, we present the exact solution for the
forcing term of the form: $F(x,t)=\omega ^2x+f(t).$ We also present the
initial value problem solution for the case with a constant forcing term to
compare with the known result.
\end{abstract}

\section{\allowbreak Introduction}

Nonlinear partial differential equations of evolution type such as the KdV,
nonlinear Schr\"{o}dinger, and sine-Gordon equations were known to be
exactly integrable in the sense that there was a transformation which mapped
these equations into a pair of linear equation that are solvable by the
Inverse Scattering Method$^1.$

These equations are of homogeneous type so the natural question arises
whether this nice integrability properties also shared by their
inhomogeneous counterpart. Several attempts had been made to extend the
Inverse Scattering Method to the inhomogeneous case, although no exact
solutions was found$^2.$

In this paper, we will be considering this problem, but instead of dealing
with the above nonlinear evolution equations, we choose the much simpler
one, namely the inhomogeneous Burgers equation,

\begin{equation}
\frac{\partial u}{\partial t}+u\frac{\partial u}{\partial x}-v\frac{\partial
^2u}{\partial x^2}=F\left( x,t\right) ,\text{ }v>0  \tag{1.1}
\end{equation}
where the inhomogeneous or source term $F\left( x,t\right) $ is a given
function. Henceforth, the $x$ and $t$ will be considered as the space and
time coordinates, respectively.

Basically, what motivated us to consider the inhomogeneous Burgers equation
(1.1) was, beside its simplicity, the well known fact that its homogeneous
counterpart is linearizable by the celebrated Hopf-Cole transformation$^3,$%
\begin{equation}
u=-v\frac 1\psi \frac{\partial \psi }{\partial x}  \tag{1.2}
\end{equation}
which enables one to solve exactly the corresponding initial value problem.

Remarkably enough, the Hopf-Cole transformation (1.2) also transforming the
inhomogeneous Burgers equation (1.1) into a linear heat type equation$%
^{4,5}, $%
\begin{equation}
\frac{\partial \psi }{\partial t}=\left[ v\frac{\partial ^2\psi }{\partial
x^2}-\frac 1{2v}V\left( x,t\right) \right] \psi  \tag{1.3$a$}
\end{equation}
where 
\begin{equation}
V\left( x,t\right) =\dint\limits_c^xdx^{\prime }F\left( x^{\prime },t\right)
+C\left( t\right)  \tag{1.3$b$}
\end{equation}
with $c$ being an arbitrary constants, while $C\left( t\right) $ an
arbitrary function of $t$.

The outline of the paper is as follows. In sec. II, we summarize the general
solution of the initial value problem of the inhomogeneous Burgers equation
(1.1), in terms of the Feynman-Kac path integral formalism. Secs. III and IV
will be dealing with the explicit exact solutions for special choices of the
source term: $F\left( x,t\right) =Q,$ a constant, and $\omega ^2x+f\left(
t\right) $, respectively. We also show that our solutions in Secs. III and
IV in the limit $\omega \rightarrow 0$, coincide with the result one would
obtain by simply applying the Orlowsky-Sobczyk transformation$^6.$

\section{Initial Value Problem and the Feynman-Kac Path Integral}

The Hopf-Cole transformation (1.2) with the linear heat equation (1.3)
enables one to apply the linear method to solve \textit{exactly }the\textit{%
\ initial value problem} (IVP) of the inhomogeneous Burgers equation (1.1).

In ref. 6, Orlowsky and Sobczyk considered the pure time dependent forcing
term, $F\left( x,t\right) =f\left( t\right) ,$ and found that the
transformations:

\begin{equation}
x^{\prime }=x-\phi \left( t\right) ,\text{ }t^{\prime }=t,\text{ }u^{\prime
}\left( x^{\prime },t^{\prime }\right) =u\left( x,t\right) -W\left( t\right)
\tag{2.1$a$}
\end{equation}
with

\begin{equation}
W\left( t\right) =\int\limits_0^td\tau f\left( \tau \right) ,\text{ }\phi
\left( t\right) =\int\limits_0^td\tau W\left( \tau \right)   \tag{2.1$b$}
\end{equation}
transform the corresponding inhomogeneous Burgers equation (1.1) to the
homogeneous form,

\begin{equation}
\frac{\partial u^{\prime }}{\partial t^{\prime }}+u^{\prime }\frac{\partial
u^{\prime }}{\partial x^{\prime }}-v\frac{\partial ^2u^{\prime }}{\partial
x^{\prime 2}}=0  \tag{2.2}
\end{equation}
Thus, the corresponding initial value problem is trivially solved by
applying the transformation (2.1).

Ablowitz and De Lillo then proceeded to consider the forcing term of the
form $F\left( x,t\right) =f\left( t\right) \delta \left( x\right) $ and
solve the corresponding IVP via a linear Voltera integral equation$^7.$

In this section, we will formulate the IVP solution for a general source
term, $F\left( x,t\right) $, via the Feynman-Kac path integral method$^8$
which proceeds as follows.

Let $u\left( x,0\right) =u\left( x\right) $ be the prescribed initial value
of the inhomogeneous Burgers equation (1.1). Then according to the Hopf-Cole
transformation (1.2), the corresponding initial value for the linear
equation (1.3) is, 
\begin{equation}
\psi \left( y,0\right) =\exp \left[ -\frac 1{2v}\tint\limits_0^ydx^{\prime
}u^{\prime }\left( x^{\prime }\right) \right]   \tag{2.3}
\end{equation}
The solution of the linear equation (1.3) is well-known, namely, 
\begin{equation}
\psi \left( x,t\right) =\int\limits_{-\infty }^\infty dy\text{ }K\left(
x,t,y,0\right) \psi \left( y,0\right)   \tag{2.4}
\end{equation}
where the kernel $K\left( x,t,y,0\right) $ satisfies the heat kernel
equation, 
\begin{equation}
\frac{\partial K}{\partial t}-v\frac{\partial ^2K}{\partial x^2}+\frac
1{2v}V\left( x,t\right) K=0  \tag{2.5}
\end{equation}
with the initial condition, $K\left( x,0,y,0\right) =\delta \left(
x-y\right) $.

Instead of solving directly the heat kernel equation (2.5), we adopt the
Feynman-Kac formulation, for which the solution is given in terms of the 
\textit{path integral} formula: 
\begin{equation}
K\left( x,t,y,0\right) =\int \left[ Dx\right] \exp \left[ -S/2v\right]  
\tag{2.6}
\end{equation}
where $S$ is the related action, \textit{i.e.}, 
\begin{equation}
S\left[ x\left( t\right) \right] =\int\limits_0^td\tau \left[ \frac 12\left( 
\frac{dx}{d\tau }\right) ^2+V\left( x,\tau \right) \right]   \tag{2.7}
\end{equation}
Following Feynman$^8$, we define the path integral (2.6) as, 
\begin{equation}
K\left( x,t,y,0\right) =\lim\Sb N\rightarrow \infty  \\ \varepsilon
\rightarrow 0 \endSb \left( \frac 1{4\pi v\varepsilon }\right)
^{N/2}\int\limits_{-\infty }^\infty dx_1\ldots dx_N\exp \left[ -S\left(
x_1,\ldots ,x_N\right) /2v\right]   \tag{2.8}
\end{equation}
where $\varepsilon =t_n-t_{n-1},$ $(n=1,2,...,N),$ $t_0=0,$ $%
t_N=N\varepsilon =t,$ $x_n=x\left( t_n\right) ,$ $y=x_0,$ $x=x_N,$ and 
\begin{equation}
S\left( x,\ldots ,y,t\right) =\frac \varepsilon 2\sum\limits_{n=1}^N\left\{
\left( \frac{x_n-x_{n-1}}\varepsilon \right) ^2+V\left( \frac{x_n+x_{n-1}}%
2,t_{n-1}\right) \right\}   \tag{2.9}
\end{equation}
The choice (2.7) for the action $S$ is verified by substituting eq. (2.9)
into (2.8) and considering the small $t=\varepsilon $ expansion, in (2.4),
to rederive the heat equation (1.3)$^8$.

\section{Solutions For a Constant Inhomogeneous Term}

To illustrate the IVP solution generating method that we presented in Sec.
II, in this section we choose to solve the inhomogeneous Burgers equation
(1.1) for the simplest case where the source term $F\left( x,t\right) $ is
chosen to be a constant, $i.e$., $F\left( x,t\right) =Q$. For this case, the
corresponding linear equation (1.3) is,

\begin{equation}
\frac{\partial \psi }{\partial t}=v\frac{\partial ^2}{\partial x^2}-\frac
1{2v}Qx\psi  \tag{3.1}
\end{equation}
where in (1.3$b$) we have chosen the lower integral limit, $c=0$, and $%
C\left( t\right) =0$. Throughout this paper we will keep to choose these
values of $c$ and $C\left( t\right) $.

In the following we apply the Feynman-Kac method to solve the corresponding
initial value problem, while the corresponding steady state solution is
presented in the Appendix A.

From eq. (1.3$b$) we find that the corresponding action (2.7) is, 
\begin{equation}
S\left[ \left( x,t\right) \right] =\int\limits_0^td\tau \left[ \frac
12\left( \frac{dx}{d\tau }\right) ^2+Qx\right]  \tag{3.2}
\end{equation}
Since the action (3.2) is quadratic, we may simplify the evaluation of the
heat kernel path integral (2.6) by expanding the action (3.2) around the
classical path $x_{cl}$, $i.e.,$

\begin{equation}
x\left( \tau \right) =x_{cl}\left( \tau \right) +\eta \left( \tau \right) 
\tag{3.3}
\end{equation}
where $x_{cl}$ obeys the Euler-Lagrange equation,

\begin{equation}
\frac{d^2x_{cl}}{d\tau ^2}=Q  \tag{3.4}
\end{equation}
By a straightforward calculation, we obtain$^5,$

\begin{equation}
K\left( x,t,y,0\right) =\left( \frac 1{4\pi vt}\right) ^{1/2}\exp \left[
-S_{cl}/2v\right]  \tag{3.5$a$}
\end{equation}
where $S_{cl}$ is the classical action

\begin{equation}
S_{cl}/2v=\xi \left( x,y,t\right) ^2+\zeta \left( x,t\right)  \tag{3.5$b$}
\end{equation}
with 
\begin{equation}
\xi \left( x,y,t\right) =\frac{2\left( x-y\right) -Qt^2}{4\sqrt{vt}},\text{ }%
\zeta \left( x,t\right) =\frac{Qt\left( 6x-Qt^2\right) }{12v}  \tag{3.5$c$}
\end{equation}

For an illustration, let us consider the single hump initial condition$^3$,

\begin{equation}
u\left( x\right) =R\delta \left( x\right)   \tag{3.6}
\end{equation}
where $\delta \left( x\right) $ is a delta function and $R$ is a constant.
The related initial condition for the linear equation (3.1), according to
(2.3) is,

\begin{equation}
\psi \left( x,0\right) =\exp \left[ -\frac 1{2v}R\eta \left( x\right)
\right]   \tag{3.7$a$}
\end{equation}
where 
\begin{equation}
\eta \left( x\right) =\lim_{\varepsilon \rightarrow
0}\int\limits_\varepsilon ^xdx\delta \left( x\right) =\left\{ _{-1,x<0}^{0,%
\text{ }x>0}\right\}   \tag{3.7$b$}
\end{equation}
with $\varepsilon >0^3$. Substituting (3.5) and (3.7) into (2.4), then the
corresponding initial value problem solution of the linear heat equation
(3.1) is,

\begin{equation}
\psi \left( x,t\right) =-\frac{\sqrt{\pi }}2e^{-\zeta \left( x,t\right)
}\left[ 2+\left( e^{R/2v}-1\right) erfc\left( \xi _0\right) \right] 
\tag{3.8$a$}
\end{equation}
where $\xi _0=\xi \left( x,0,t\right) ,$ and 
\begin{equation}
erfc\left( z\right) =\frac 2{\sqrt{\pi }}\int\limits_z^\infty d\xi e^{-\xi
^2}  \tag{3.8$b$}
\end{equation}
is the complementary error function$^9$.

Substituting (3.8) into the Hopf-Cole transformation (1.2) gives the
required solution of the inhomogeneous Burgers equation (1.1) with $F\left(
x,t\right) =Q$, $i.e.$, 
\begin{equation}
u\left( x,t\right) =Qt+\sqrt{\frac{4v}{\pi t}}\frac{\left(
e^{-R/2v}-1\right) e^{-\xi _0^2}}{2+\left( e^{-R/2v}-1\right) erfc\left( \xi
_0\right) }  \tag{3.9}
\end{equation}
which obviously reduces to the corresponding homogeneous solution for $Q=0^3$%
.

Starting from the corresponding homogeneous solution with the single hump
initial condition (3.6)$^3$, one rederives the solution (3.9) directly by
applying the Orlowsky-Sobczyk transformation (2.1).

\section{Solutions for $F\left( x,t\right) =\omega ^2x+f\left( t\right) $}

In this section we give an explicit IVP solution for the forcing term $%
F\left( x,t\right) =\omega ^2x+f\left( t\right) $, for which the action
(2.7) is still of quadratic type, $i.e.$, 
\begin{equation}
S\left[ \left( x,t\right) \right] =\int\limits_0^td\tau \left[ \frac
12\left( \frac{dx}{d\tau }\right) ^2+\frac 12\omega ^2x^2+f\left( \tau
\right) x\right]   \tag{4.1}
\end{equation}
where $\omega $ is a constant, and $f\left( t\right) $ a given function of
time, $t$.

Applying the steepest descent method$^8$ to compute the heat-Kernel $K\left(
x,t,y,0\right) $ in (2.6) for a quadratic action, as in Sec. III, we obtain
the exact kernel: 
\begin{equation}
K\left( x,t,y,0\right) =N\exp \left[ -S_{cl}/2v\right]  \tag{4.2}
\end{equation}
where $S_{cl}$ is the classical action (see the Appendix B): 
\begin{equation}
\begin{array}{c}
S_{cl}=\frac \omega {2\sinh \omega t}\left[ \left( x^2+y^2\right) \cosh
\omega t-2xy\right] \\ 
+\frac 1{\sinh \omega t}\left[ x\int\limits_0^td\tau \sinh \left( \omega
\tau \right) f\left( \tau \right) +y\int\limits_0^td\tau \sinh \omega \left(
t-\tau \right) f\left( \tau \right) \right] \\ 
-\frac 1{\omega \sinh \omega t}\left[ x\int\limits_0^td\tau
\int\limits_0^\tau d\tau ^{\prime }\sinh \omega \left( t-\tau \right) \sinh
\left( \omega \tau ^{\prime }\right) f\left( \tau ^{\prime }\right) f\left(
\tau \right) \right]
\end{array}
\tag{4.3}
\end{equation}
and $N$ the corresponding prefactor which is independent of $x$ and $y$.

Thus, the corresponding IVP solution for $u\left( x,t\right) $, according to
the Hopf-Cole transformation (1.2), with $\psi \left( x,t\right) $ given by
(2.4), after substituting the heat-Kernel (4.2), is: 
\begin{equation}
\begin{array}{c}
u\left( x,t\right) =\left( 2Ax+C\right)  \\ 
-\frac \omega {\sinh \omega t}\frac{\int\limits_{-\infty }^\infty dy\text{ }%
y\exp \left[ -A\left( y+B/2A\right) ^2/2v\right] \psi \left( y,0\right) }{%
\int\limits_{-\infty }^\infty dy\exp \left[ -A\left( y+B/2A\right)
^2/2v\right] }
\end{array}
\tag{4.4$a$}
\end{equation}
where 
\begin{equation}
A=\frac 12\omega \coth \omega t  \tag{4.4$b$}
\end{equation}
\begin{equation}
B=-\frac \omega {\sinh \omega \tau }x+\frac 1{\sinh \omega
t}\int\limits_0^td\tau \sinh \omega \left( t-\tau \right) f\left( \tau
\right)   \tag{4.4$c$}
\end{equation}
\begin{equation}
C=\frac 1{\sinh \omega t}\left[ \int\limits_0^td\tau \sinh \omega t\text{ }%
f\left( \tau \right) \right]   \tag{4.4$d$}
\end{equation}

For an illustration, let us consider the shock profile initial condition$^1$%
, 
\begin{equation}
u\left( x,0\right) =(\nu a)[1-\tanh (ax/2)]  \tag{4.5}
\end{equation}
with $a$ a constant, for which the corresponding initial condition for $\psi
(x,t)$, according to (2.3) is, 
\begin{equation}
\psi (x,0)=2[1+e^{-ax}]  \tag{4.6}
\end{equation}
Substituting (4.6) into (4.4), then by a straightforward manipulations, we
obtain the corresponding initial value problem solution: 
\begin{equation}
u\left( x,t\right) =\left( 2Ax+C\right) -\frac \omega {2A\sinh \omega
t}\left[ B+\frac{2vae^{(aB+va^2)/A}}{1+e^{(aB+va^2)/A}}\right]  \tag{4.7}
\end{equation}
which reduces to the celebrated Taylor shock profile solution$^1$ in the
limit: $\omega \rightarrow 0,$ and $f\left( t\right) \rightarrow 0$.

Taking instead the limit $\omega \rightarrow 0$, but $f(t)\neq 0$, we obtain
from (4.7), the reduced solution, 
\begin{equation}
u(x,t)=W(t)+(\nu a)\left[ 1-\tanh \left( \left\{ a(x+\phi (t)-at\right\}
/2\right) \right]  \tag{4.8}
\end{equation}
where $\phi (t)$ and $W(t)$ are given by (2.1$b$), as one would expect by
applying the Orlowsky-Sobczyk transformation (2.1) to the IVP solution of
the corresponding homogeneous Burgers equation$^{1,6}$. It should be
mentioned that in obtaining the solution (4.8), we have used the following
limiting values of $A,$ $B,$ and $C$, obtainable from (4.4), $i.e,$%
\begin{equation}
A=\frac 1{2t}  \tag{4.9$a$}
\end{equation}
\begin{equation}
B=-\frac xt+\frac 1t\int\limits_0^td\tau (t-\tau )f(\tau )=-\frac xt+\phi (t)
\tag{4.9$b$}
\end{equation}
\begin{equation}
C=\frac 1t\left[ \int\limits_0^td\tau f(\tau )\right]  \tag{4.9$c$}
\end{equation}
where in (4.9$b$), we use the fact that: 
\begin{equation}
\phi (t)=\int\limits_0^td\tau W(\tau )=\int\limits_0^td\tau
\int\limits_0^\tau d\tau ^{\prime }f(\tau ^{\prime })=\int\limits_0^td\tau
^{\prime }(t-\tau ^{\prime })f(\tau ^{\prime })  \tag{4.10}
\end{equation}

\section{\textbf{Acknowledgements}}

It is a pleasure for us to acknowledge the Directorate General of Higher
Education of the Republic of Indonesia, for supporting this research under
the project: Hibah Bersaing V, 1996-1998. We also thank Drs. M. Husin Alatas
and Bobby E. Gunara for discussion, and Prof. P. Silaban for the
encouragement.

\section*{\textbf{Appendix A. The steady state solution for }$F\left(
x,t\right) =Q$}

In the following, we present the steady state solution of the heat equation
(3.1). First of all, let us look for a single state solution, for which $%
\psi \left( x,t\right) $ is given by, 
\begin{equation}
\psi \left( x,t\right) =T\left( t\right) \varphi \left( x\right)   \tag{A.1}
\end{equation}
Substituting (A.1) into (3.1) gives the separated equations: 
\begin{equation}
\frac{dT}{dt}=\omega T  \tag{A.2}
\end{equation}
\begin{equation}
\frac{d^2\varphi }{dx^2}-z\varphi =0  \tag{A.3}
\end{equation}
where 
\begin{equation}
z=\frac xl+\lambda \text{, }\lambda =\left( \frac \omega v\right) l^2,\text{ 
}l=\left( \frac{2v^2}Q\right) ^{1/3}  \tag{A.4}
\end{equation}
and $\omega $ is the corresponding separation constant.

The solution of eq. (A.3) is well-known, i.e.: 
\begin{equation}
\varphi \left( x\right) =\left[ aAi\left( z\right) +bBi\left( z\right)
\right]  \tag{A.5}
\end{equation}
where $Ai\left( z\right) $ and $Bi\left( z\right) $ are the Airy functions
of the first and second kind, respectively$^9$, with $a$ and $b$ are
constants which are fixed by the boundary conditions.

Thus, the single state solution of eq. (3.1) is given by 
\begin{equation}
\psi \left( x,t\right) =e^{\omega t}\left[ aAi\left( z\right) +bBi\left(
z\right) \right]   \tag{A.6}
\end{equation}
Unfortunately, the corresponding solution for $u\left( x,t\right) $ under
the Hopf-Cole transformation (1.2), is a \textit{static }solution.

However, since eq. (3.1) is linear, we may take the superposition, 
\begin{equation}
\psi \left( x,t\right) =\sum\limits_{n=1}^{N>1}e^{\omega _nt}\left[
a_nAi\left( z_n\right) +b_nBi\left( z_n\right) \right]   \tag{A.7$a$}
\end{equation}
where 
\begin{equation}
z_n=\frac xl+\lambda _n,\text{ }\lambda _n=\left( \frac{\omega _n}v\right)
l^2  \tag{A.7$b$}
\end{equation}
as the solution for eq. (3.1). The corresponding solution for the
inhomogeneous Burgers equation (1.1), with $F\left( x,t\right) =Q$, through
the Hopf-Cole transformation (1.2) is, 
\begin{equation}
u\left( x,t\right) =-\frac{2v}l\frac{\sum\limits_{n=1}^{N>1}e^{\omega
_nt}\left[ a_nAi^{\prime }\left( z_n\right) +b_nBi^{\prime }\left(
z_n\right) \right] }{\sum\limits_{n=1}^{N>1}e^{\omega _nt}\left[ a_nAi\left(
z_n\right) +b_nBi\left( z_n\right) \right] }  \tag{A.8}
\end{equation}
with $Ai^{\prime }\left( z_n\right) $ and $Bi^{\prime }\left( z_n\right) $
are the derivatives of the corresponding Airy functions with respect to $z_n$%
. The solution (A.8) is obviously \textit{non-static.}

\section*{\textbf{Appendix B. Derivation of the classical action (4.3)}}

In this Appendix, we present the derivation of the classical action (4.3).
For this purpose, let us rewrite the action (4.1) in the following form: 
\begin{equation}
2S=x\left( \tau \right) \frac{dx}{d\tau }\mid
_{t=t_i}^{t=t_f}-\int\limits_{t_i}^{t_f}d\tau \left\{ x\frac{d^2x}{d\tau ^2}%
-\frac 12\omega ^2x^2-2f\left( \tau \right) x\right\}  \tag{B.1}
\end{equation}
Then, by using the equation of motion: 
\begin{equation}
\frac{d^2x}{d\tau ^2}-\omega ^2x-f\left( \tau \right) =0  \tag{B.2}
\end{equation}
the corresponding classical action is, 
\begin{equation}
2S=x\left( \tau \right) \frac{dx}{d\tau }|_{t=t_i}^{t=t_f}-\int%
\limits_{t_i}^{t_f}d\tau \text{ }x\left( \tau \right) f\left( \tau \right) 
\tag{B.3}
\end{equation}
where $x\left( \tau \right) $ is given by the solution of (B.2), with the
boundary conditions, $x\left( t_i\right) =x_i$ and $x\left( t_f\right) =x_f$.

Using the Green's function method, the solution of (B.2) is, 
\begin{equation}
x\left( \tau \right) =\int\limits_{t_i}^{t_f}d\tau ^{\prime }G\left( \tau
^{\prime },\tau \right) f\left( \tau \right) +x\left( \tau ^{\prime }\right)
\frac d{d\tau ^{\prime }}G\left( \tau ^{\prime },\tau \right) \mid
_{t=t_i}^{t=t_f}  \tag{B.4}
\end{equation}
where the Green's function G satisfies: 
\begin{equation}
\frac{d^2}{d\tau ^2}G\left( \tau ^{\prime },\tau \right) +\omega ^2G\left(
\tau ^{\prime },\tau \right) =\delta \left( \tau ^{\prime }-\tau \right) 
\tag{B.5}
\end{equation}
with the boundary conditions, 
\begin{equation}
G\left( \tau ^{\prime },t_i\right) =G\left( \tau ^{\prime },t_f\right) =0 
\tag{B.6}
\end{equation}
Solving eq. (B.5) for $\tau <\tau ^{\prime }$ and $\tau >\tau ^{\prime }$,
then using the boundary conditions (B.6), the continuity condition, 
\begin{equation}
G\left( \tau ^{\prime }+\varepsilon ,\tau ^{\prime }\right) -G\left( \tau
^{\prime }-\varepsilon ,\tau ^{\prime }\right) =0,\text{ }\left( \varepsilon
\rightarrow 0\right)  \tag{B.7}
\end{equation}
and the discontinuity condition, 
\begin{equation}
\frac d{d\tau }G\left( \tau ,\tau ^{\prime }\right) \mid _{\tau =\tau
^{\prime }+\varepsilon }-\frac d{d\tau }G\left( \tau ,\tau ^{\prime }\right)
\mid _{\tau =\tau ^{\prime }-\varepsilon }=0,\text{ }\left( \varepsilon
\rightarrow 0\right)  \tag{B.8}
\end{equation}
we obtain: 
\begin{equation}
G\left( \tau ,\tau ^{\prime }\right) =\frac 1{\omega \sinh \omega \left(
\tau _f-\tau _i\right) }\left\{ 
\begin{array}{c}
\sinh \omega \left( \tau ^{\prime }-\tau _f\right) \sinh \omega \left( \tau
-\tau _i\right) ,\text{ }\tau <\tau ^{\prime } \\ 
\sinh \omega \left( \tau ^{\prime }-\tau _i\right) \sinh \omega \left( \tau
-\tau _f\right) ,\text{ }\tau >\tau ^{\prime }
\end{array}
\right\}  \tag{B.9}
\end{equation}
Substituting (B.9) in (B.4) gives: 
\begin{equation}
\begin{array}{c}
x\left( t\right) =\frac 1{\sinh \omega \left( \tau _f-\tau _i\right) }\left[
x_f\sinh \omega \left( \tau -\tau _i\right) -x_i\sinh \omega \left( \tau
-\tau _f\right) \right] \\ 
-\frac 1\omega \int\limits_{t_i}^\tau d\tau ^{\prime }\sinh \omega \left(
\tau -\tau _f\right) \sinh \omega \left( \tau ^{\prime }-\tau _i\right)
f\left( \tau ^{\prime }\right) \\ 
-\frac 1\omega \int\limits_\tau ^{t_f}d\tau ^{\prime }\sinh \omega \left(
\tau -\tau _i\right) \sinh \omega \left( \tau ^{\prime }-\tau _f\right)
f\left( \tau ^{\prime }\right)
\end{array}
\tag{B.10}
\end{equation}
Using (B.10) in (B.3), and choose :$\tau _i=0,$ $\tau _f=t,$ $x_i=y,$ $%
x_f=x, $ then by a straightforward manipulations we obtain the classical
action $S_{cl}$ in (4.3).

\end{document}